\documentclass{raa}
\usepackage{graphicx,times}          
\usepackage{natbib}
\usepackage{amssymb,amsmath}
\usepackage{float}
\usepackage{upgreek}                      
\usepackage{longtable}
\usepackage{ltxtable}
\usepackage{placeins}
\usepackage{tabularx} 
\usepackage{tablefootnote}
\usepackage{graphicx}

\bibpunct{(}{)}{;}{a}{}{,}

\usepackage[pagebackref=true]{hyperref}

\begin{document}

  \title{Identifying symbiotic stars with machine learning
}

   \volnopage{Vol.0 (20xx) No.0, 000--000}      
   \setcounter{page}{1}          

   \author{Yongle Jia
      \inst{1}
   \and Sufen Guo 
      \inst{1*}
   \and Chunhua Zhu 
      \inst{1}
   \and Lin Li 
      \inst{1}
   \and Mei Ma 
      \inst{1}
   \and Guoliang Lü 
      \inst{2,1*}
   }

   \institute{School of Physical Science and Technology, Xinjiang University,
               Urumqi 830046, China; {\it guosufen@xju.edu.cn}\\
        \and
        Xinjiang Astronomical Observatory, Chinese Academy of Sciences, 150 Science 1-Street,
              Urumqi, Xinjiang 830011, China; {\it guolianglv@xao.ac.cn}\\
\vs\no
   {\small Received 2023 ****; accepted 2023 ****}}

\abstract
{Symbiotic stars are interacting binary systems, making them valuable for studying various astronomical phenomena, such as stellar evolution, mass transfer, and accretion processes. Despite recent progress in the discovery of symbiotic stars, a significant discrepancy between the observed population of symbiotic stars and the number predicted by theoretical models. To bridge this gap, this study utilized machine learning techniques to efficiently identify new symbiotic stars candidates. Three algorithms (XGBoost, LightGBM, and Decision Tree) were applied to a dataset of 198 confirmed symbiotic stars and the resulting model was then used to analyze data from the LAMOST survey, leading to the identification of 11,709 potential symbiotic stars candidates. Out of the these potential symbiotic stars candidates listed in the catalog, 15 have spectra available in the SDSS survey. Among these 15 candidates, two candidates, namely V* V603 Ori and V* GN Tau, have been confirmed as symbiotic stars. The remaining 11 candidates have been classified as accreting-only symbiotic star candidates. The other two candidates, one of which has been identified as a galaxy by both SDSS and LAMOST surveys, and the other identified as a quasar by SDSS survey and as a galaxy by LAMOST survey.
}

   \authorrunning{Y.-L. Jia, G.-F. Guo, Z.-C. Hua, L.-Li, M. Ma  \& G.-L. Lü }         
   \titlerunning{Identifying symbiotic stars with machine learning}  

   \maketitle
\section{Introduction}     
\label{sect:intro} 
Symbiotic stars are a unique type of binary system, characterized by prolonged interactions between the two stars (\citealt{2006MNRAS.372.1389L,2009MNRAS.396.1086L,2012MNRAS.424.2265L}; \citealt{2020RAA....20..161H}). 
These systems are composed of three components: a hot companion such as a white dwarf, neutron star, or main sequence star with an accretion disk; a cool companion, 
such as a red giant or asymptotic giant branch star; and an ionized nebula formed from material lost by the cool companion (\citealt{1991AJ....101..637K}; \citealt{2021MNRAS.505.6121M}). 
The spectrum of symbiotic stars is composed of common features, including emission lines from the hot companion, absorption lines from the cool companion, 
and the nebula's material. In many symbiotic stars, the cool companion loses mass through stellar wind or Roche-lobe overflow, while the hot companion accretes enough material to produce the symbiotic phenomenon (\citealt{2007BaltA..16....1M}; \citealt{2021MNRAS.502.2513A}; \citealt{2022MNRAS.510.2707I}). 
Symbiotic stars provide an excellent sample for studying the loss of matter, acceleration mechanisms of stellar winds, and accretion of stellar winds in late-type giants (\citealt{2017MNRAS.468.4465C}; \citealt{2019A&A...626A..68S}). 
They are also considered to be the precursors of $\rm\uppercase\expandafter{\romannumeral1}$a supernovae and important sources of soft and hard X-rays (\citealt{1992ApJ...397L..87M}).

Since symbiotic stars are unique astrophysical laboratories.
\citet{1984PASA....5..369A} provided a catalog of symbiotic stars, including 129 symbiotic stars and 15 symbiotic stars candidates. 
\citet{2009syst.book.....K} summarized a catalog of symbiotic stars, including 133 symbiotic stars and 20 symbiotic stars candidates. 
\citet{2000A&AS..146..407B} provided a more detailed catalog of 188 symbiotic stars and 30 symbiotic stars candidates. 
\citet{2019ApJS..240...21A} provided a catalog of 323 known symbiotic stars, 257 are Galactic and 66 extragalactic. 
\citet{2023MNRAS.519.6044A} obtained 814 symbiotic star candidates by GALEX UV and 2MASS/AllWISE IR photometry, and identified two symbiotic stars after spectral analysis.
\citet{2019ApJS..240...21A} proposed a new subfamily of symbiotic stars for the first time: the S+IR type.

The symbiotic stars are classified into four categories: S, D, D', and S+IR (\citealt{2019ApJS..240...21A}). 
\citet{2019ApJS..240...21A} used a blackbody radiation model to fit the spectral energy map of symbiotic stars and found a number of S-type symbiotic stars with an S-type SED profile and an infrared excess in the mid-infrared regime. 
\citet{2019ApJS..240...21A} named these symbiotic stars S+IR type. 
Most of the symbiotic stars are S-type, and the SED profiles of S-type symbiotic stars peak at 0.8 to 1.7 $\upmu$m with an average value of 1.07 $\upmu$m. Of course, a small percentage of S-type symbiotic stars with SED diagrams peaking around 0.7 or 1.8 $\upmu$m. The temperature is between 3400 and 3800 K. Thirteen percent of the symbiotic stars are D-type, and the SED diagram for D-type symbiotic stars peaks at 2 to 4 $\upmu$m with an average value of 2.85 $\upmu$m. The number of D'-type symbiotic stars is relatively small, with only 10 known symbiotic stars with G/K spectral type. The SED diagram of D'-type symbiotic stars has two peaks between 2 and 10 $\upmu$m. The SED diagram of S+IR-type symbiotic stars has a peak at 1.3 $\upmu$m. The IR excess indicates the presence of a dusty shell around the symbiotic star and a lower temperature than that of the D-type. It is most likely that there is an accretion disk around the white dwarf (\citealt{2019ApJS..240...21A}). Currently, there is a lack of in-depth research on the spectral characteristics of S+IR-type symbiotic stars. However, \citet{ 1996PASP..108..972G, 1998A&A...333..658P, 1999A&AS..137..473M, 2005A&A...429..993P, 2007ASSL..342.....K, 2020MNRAS.495.1461S} had conducted research on the spectral characteristics of S-type, D-type and D'-type symbiotic stars.
The latest list of symbiotic stars was summarized by \citet{2020CoSka..50..426M}, with 275 symbiotic stars and 119 symbiotic stars candidates in the galaxy. 
Although the number of observed symbiotic stars is increasing, the number of observed symbiotic stars differs significantly from what was predicted. 
\citet{2006MNRAS.372.1389L} predicted that the number of symbiotic stars with white dwarf(WD) accretors in the Galaxy may range from about 1200 to 15000, and the model birth rate of symbiotic stars in the Galaxy is from 0.035-0.131 yr$^{-1}$. 
It is predicted that the number of symbiotic stars in the galaxy should be from 3$\times$10$^{3}$ to 4$\times$10$^{5}$ (\citealt{1984PASA....5..369A}; \citealt{2003ASPC..303..539M}; \citealt{2006MNRAS.372.1389L}). 
The number of known symbiotic stars is very different from the predicted number, so we need to discover more symbiotic stars to provide astronomers with data. 
Previously, symbiotic stars were found by spectroscopic analysis. With the introduction of various telescopes, astronomical data has grown exponentially, 
and the previous methods of processing data are unsuitable for the current requirements of processing large data. 
We need a new method to discover new symbiotic stars from the huge amount of astronomical observation data.

Machine learning has become a well-established tool in the field of astronomy. 
For instance, \citet{1998ESASP.413..711G} and \citet{1998ASPC..138..309S} employed artificial neural networks to classify stellar spectra, 
while \citet{2014NewA...28...35B} applied support vector machines to achieve a more detailed classification of stellar spectra. 
\citet{2018A&A...618A.144G} utilized support vector machines to establish Wide-field Infrared Survey Explorer (WISE) mid-infrared color criteria for the selection of quasar candidates, resulting in the compilation of a catalog of quasar candidates. 
\citet{2021ApJS..254....6F} employed the Extreme Gradient Boosting (XGBoost) algorithm for machine learning to Pan-STARRSI (PSI) and AllWISE photometry in order to classify quasars behind the Galactic plane (GPQ). 
This led to the development of a reliable GPQ candidate catalog. 
\citet{2021MNRAS.502.2513A} utilized machine learning methods to discover five new symbiotic stars and proposed a novel method for their identification, thus rendering the determination of symbiotic stars no longer dependent on spectroscopic analysis alone. 
However, the method employed by Akras only distinguished symbiotic stars from other objects containing $\rm H$ emission lines. It is likely that other types of objects exist that have not been considered. Therefore, 
it is imperative to develop a model that can be applied more broadly. In this study, we expand upon this idea and propose a method that can be used to quickly identify symbiotic stars among a large number of objects. 
We used the aggregated coordinates of symbiotic stars that were cross-matched with AllWISE and Two Micron All Sky Survey (2MASS) data for machine learning training, and then applied the trained machine model to identify new symbiotic stars in Large Sky Area Multi-Object Fiber Spectroscopic Telescope (LAMOST). 
Through the application of machine learning, we were able to identify a new set of symbiotic stars candidates. 

The paper is divided into several sections. in Section 2, we provide a detailed description of the data sources utilized, including AllWISE, 2MASS, LAMOST, and Sloan Digital Sky Survey (SDSS), and the composition of the training data. In Section 3, we discuss the machine learning models selected and the training process. In Section 4, we present our prediction results and the spectral analysis of two newly-discovered symbiotic stars. Lastly, in Section 5 we summarize our results.

\section{Data}
\label{sect:Data}

\subsection{The AllWISE catalog}

The WISE is a medium class explorer mission funded by NASA (\citealt{2004SPIE.5487..101D}; \citealt{2010AJ....140.1868W}; \citealt{2008SPIE.7017E..0ML}).
WISE uses a 40 cm telescope to image the entire sky in four infrared bands W1, W2, W3, and W4 at 3.4, 4.6, 12, and 22 $\upmu$m
and has already produced over a million images and hundreds of millions of celestial bodies have been observed. 
The AllWISE catalog (\citealt{2014yCat.2328....0C}) extends the results of the Wide-field Infrared Survey Mission (\citealt{2012yCat.2311....0C}).
It combines data from the cryogenic and post-cryogenic periods to provide the most comprehensive mid-infrared overview currently available.
The AllWISE Source Catalog contains accurate positions, motion measurements, photometry and ancillary information for 747,634,026 objects (\citealt{2013wise.rept....1C}).
The AllWISE survey has yielded better photometric sensitivity, accuracy, and astrometric precision data than the WISE survey.

\subsection{The 2MASS catalog}

The 2MASS observes the sky in the near-infrared J (1.25 $\upmu$m ), H (1.65 $\upmu$m), and Ks (2.16 $\upmu$m) band separately, covering 99.998$\%$ of the sky observed from both the northern 2MASS facility at Mt. Hopkins, AZ, and the southern 2MASS facility at Cerro Tololo, Chile(\citealt{2003tmc..book.....C}; ). 
The release data products include 4,121,439 Atlas Images in the three survey bands, and Catalogs containing positional and photometric information for 470,992,970 Point sources and 1,647,599 Extended sources(\citealt{1992ASPC...34..203K}; \citealt{1994Ap&SS.217...11K}; \citealt{2006AJ....131.1163S}).

\subsection{The LAMOST catalog}

The LAMOST is a new type of wide field of view and large aperture telescope that is a special quasi-meridian reflecting Schmidt telescope located
in Xinglong Station of the National Astronomical Observatory, China (\citealt{2012RAA....12.1197C}; \citealt{2012RAA....12..723Z}; \citealt{2015RAA....15.1095L}).
LAMOST is an international leader in the field of wide-field optical spectroscopy and astronomy. It observes astronomical spectra in the northern sky.
LAMOST began its first spectroscopic survey in September 2012 and has released its tenth data release (LAMOST DR 10 v0),
containing more than 11 million spectra of 10 million stars, 242,569 galaxies, and 76,167 quasars.

\subsection{The SDSS catalog}

The SDSS (\citealt{1998AJ....116.3040G}; \citealt{2000AJ....120.2615F}; \citealt{2000AJ....120.1579Y}) is a wide-field optical/infrared imaging and spectroscopy survey
using a dedicated 2.5-m wide-angle optical telescope at Apache Point Observatory.
SDSS started in 1998 and has completed four phases: 
$\rm\uppercase\expandafter{\romannumeral1}$, $\rm\uppercase\expandafter{\romannumeral2}$, $\rm\uppercase\expandafter{\romannumeral3}$, $\rm\uppercase\expandafter{\romannumeral4}$ and $\rm\uppercase\expandafter{\romannumeral5}$.
Currently, the SDSS 18th data (DR18) has been released (\citealt{2023arXiv230107688A}). SDSS provides images, spectra, and scientific catalogs.
To date, more than 6 million spectra have been observed by SDSS (\citealt{2022ApJS..259...35A}).

\subsection{Feature selection}

The 2MASS J-H versus H-Ks color-color distribution diagrams have been widely used to investigate the near-infrared properties of symbiotic stars, and to differentiate between S-type, D-type, or other new candidates(\citealt{1974MNRAS.167..337A}; \citealt{2014A&A...567A..49R}; \citealt{2013AJ....146..115B}). \citet{2016AJ....151..100B} discovered that the W3-W4 color index from the WISE survey can differentiate normal K giants from D-type symbiotic stars. \citet{2019MNRAS.483.5077A} also utilized the magnitudes from both 2MASS and WISE surveys to distinguish symbiotic stars from other objects with strong H emission lines. Therefore, in our study, we used magnitudes from seven bands (W1, W2, W3, W4, J, H, and Ks) as the features for model training.

\subsection{Data set composition}
\begin{figure}
  \centering
  \includegraphics[scale=0.5]{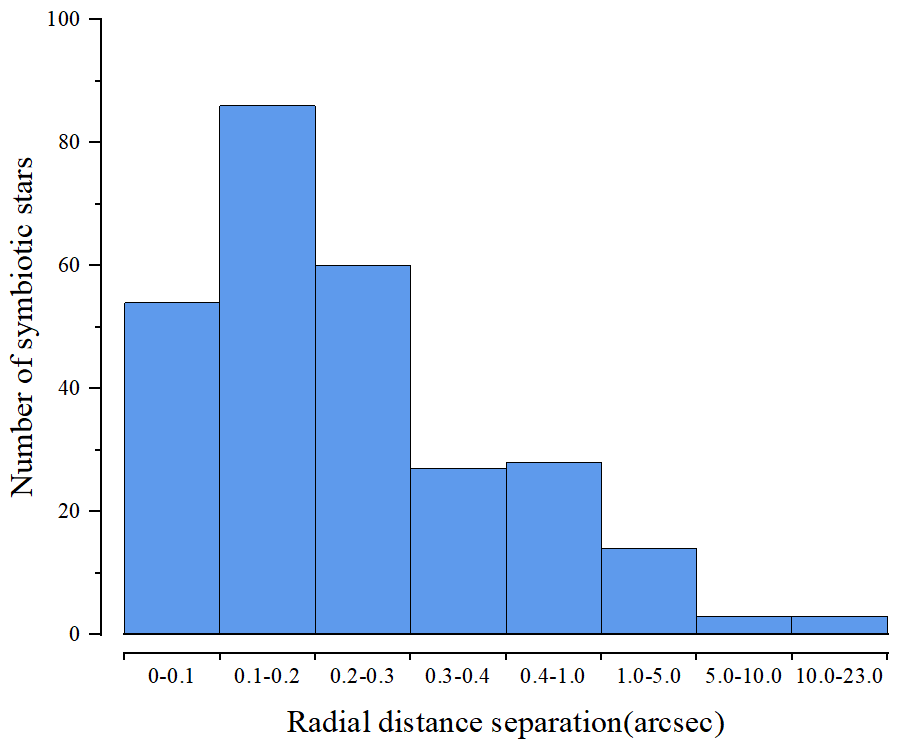}
  \caption{The radial distance between the 275 symbiotic stars and the matched sources in AllWISE. }
  \label{Fig 1}
\end{figure}

Currently, \citet{2020CoSka..50..426M} summarized the previously discovered and confirmed symbiotic stars into a new symbiotic star catalog. 
As of September 2022, this catalog contains 275 Galactic and 200 extragalactic symbiotic stars. 
This is by far the most comprehensive catalog of symbiotic stars. 
Since a lot of observational data about stars in the Milky Way, 
we only use the total of 275 Galactic symbiotic stars summarized in this catalog. 
We cross-matched the aggregated 275 symbiotic stars with AllWISE at a radius of 10 arcsec due to the angular resolution of the WISE survey, which is 6.1$^{\prime \prime}$, 6.4$^{\prime \prime}$, 6.5$^{\prime \prime}$ \& 12.0$^{\prime \prime}$ at 3.4, 4.6, 12 \& 22 $\upmu$m  (\citealt{2010AJ....140.1868W}). 
The cross-match radius between the target source and the matched source is shown in Figure \ref{Fig 1}. 
The sources in the catalog are cross-matched to their corresponding sources in AllWISE, 
and 82.5$\%$ (227) of them are cross-matched to sources in AllWISE with a radius less than 0.4 arcsec, only 3 sources were found to have a radius greater than 10 arcsec.
We believe that these 3 sources do not have corresponding sources in AllWISE. 
We performed cross-matching operations using TOPCAT (\citealt{2005ASPC..347...29T}). The AllWISE catalog has been cross-matched with 2MASS with a matching radius of 3 arcsec (\citealt{2013wise.rept....1C}). Therefore, when cross-matching a catalog with the AllWISE catalog, the resulting match will include magnitude information from the 2MASS catalog. Two symbiotic stars were missing some magnitude information, thus leaving a total of 270 symbiotic stars available for training.
We randomly divided the 270 symbiotic stars available for training into two parts. 
The first part consisted of 198 symbiotic stars used as positive samples to compose the data set with non-symbiotic stars. The second part consisted of 72 symbiotic stars used separately to test the trained models.

We used the LAMOST DR9 V1.0 spectroscopic dataset, which comprises a vast collection of 10,907,516 star spectra, 242,569 galaxy spectra, and 76,167 quasar spectra. To compose a training set, we employed a random selection strategy to choose one spectrum from every 500 star spectra, one spectrum from every 20 galaxy spectra, and one spectrum from every 3 quasar spectra. We cross-matched the selected sources with AllWISE using a matching radius of 1 arcsec to ensure accurate matches. We removed spectra without 2MASS magnitudes, resulting in a dataset of 12,348 star spectra, 9,014 galaxy spectra, and 3,797 quasar spectra. Finally, we composed the dataset by using these non-symbiotic stars as negative samples together with the 198 positive samples.

\begin{figure}
  \centering
  \includegraphics[scale=0.2]{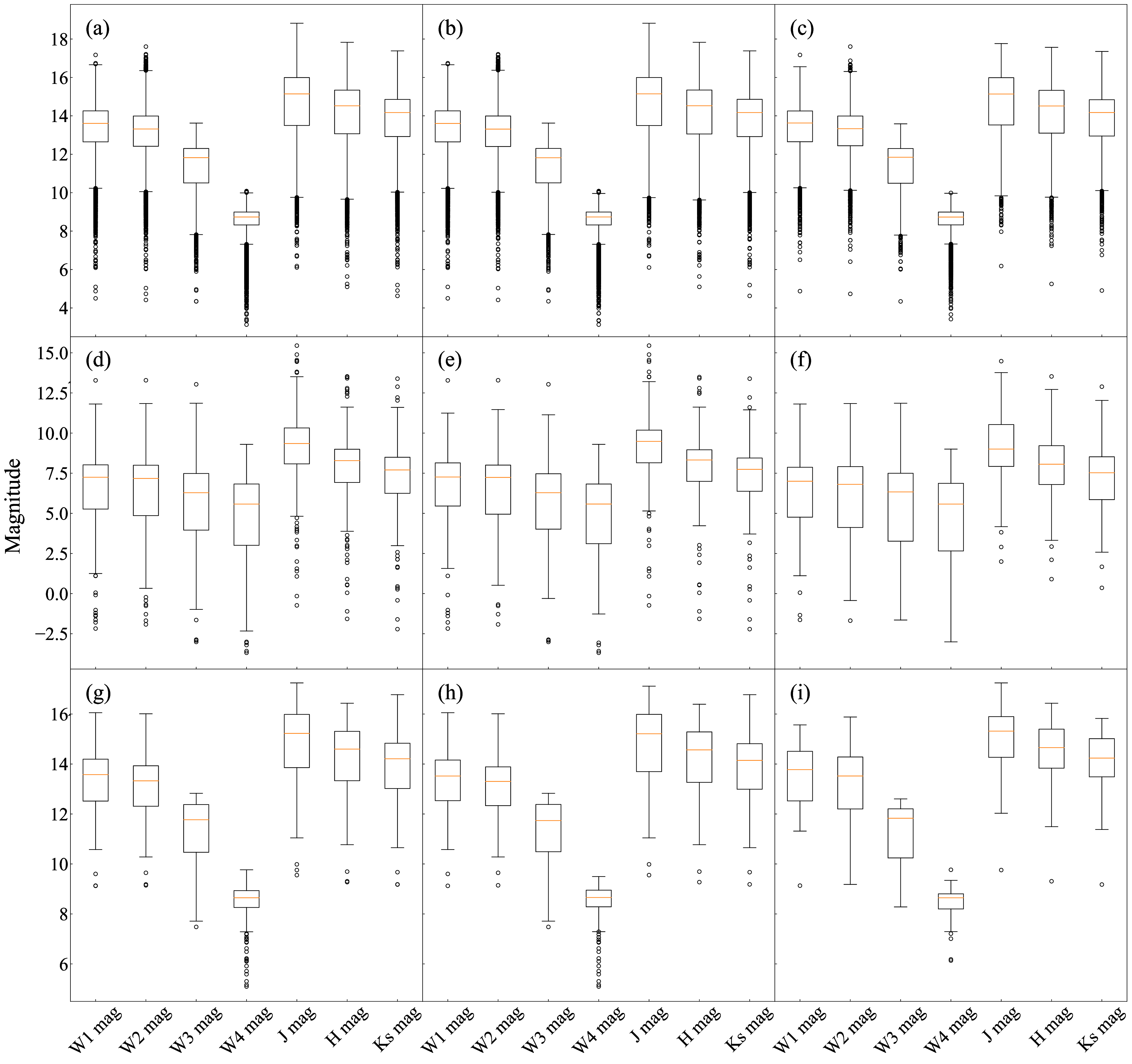}
  
  \caption{The sequence of boxplots displays the magnitude distribution of different sets of stars. Specifically, Boxplots (a) exhibit the magnitude distribution of 25,159 non-symbiotic stars, with Boxplots (b) and (c) presenting the magnitude distribution of the subset assigned to the training and testing sets, respectively. Similarly, Boxplots (d) show the magnitude distribution of 270 symbiotic stars, with Boxplots (e) and (f) displaying the magnitude distribution of the subset assigned to the training and validation sets, respectively. Lastly, Boxplots (g) illustrate the magnitude distribution of 200 randomly selected non-symbiotic stars, with Boxplots (h) and (i) depicting the magnitude distribution of the subset assigned to the training and testing sets, respectively. The horizontal orange line inside the box in the boxplot represents the median of the sample data for the corresponding band. The upper and lower limits of the box represent the upper quartile and lower quartile of the sample data for the corresponding band, respectively. There is a line above and below the box, which represents the maximum and minimum values of the sample data in the corresponding band. Hollow dots indicate some outliers in the sample data in the corresponding band.}
  \label{Fig2}
\end{figure}

Due to the significant difference in the number of symbiotic stars and non-symbiotic stars in the training set, we randomly selected an additional 200 non-symbiotic stars from the non-symbiotic star samples in the training set as additional training set samples.
Additionally, boxplots were used to depict the distribution of the training and test sets in different bands, as illustrated in Figure \ref{Fig2}. The horizontal orange line inside the box in the boxplot represents the median of the sample data for the corresponding band. The upper and lower limits of the box represent the upper quartile and lower quartile of the sample data for the corresponding band, respectively. There is a line above and below the box, which represents the maximum and minimum values of the sample data in the corresponding band. Hollow dots indicate some outliers in the sample data in the corresponding band.
As illustrated in Figure \ref{Fig2} , it can be observed that the two parts of the data exhibit a homogeneous distribution.

\begin{figure}
  \centering
  \includegraphics[scale=0.25]{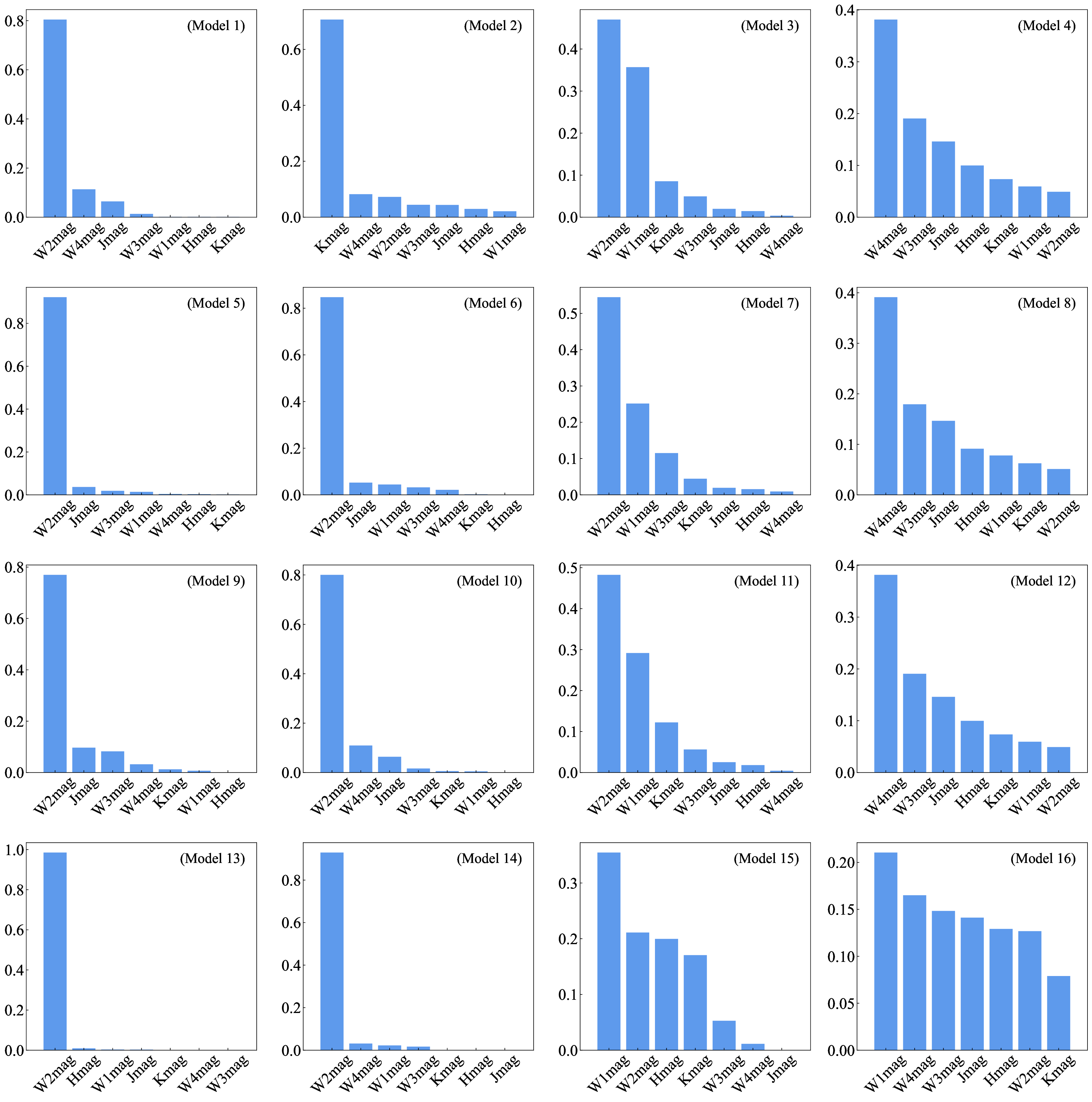}
  \caption{Distribution chart of feature importance for 16 best machine learning models.}
  \label{Fig3}
\end{figure}

\section{Model training}
\label{sect:model}

\subsection{ Model selection}
In this work, we utilized three machine learning algorithms: the Decision Tree algorithm, the XGBoost algorithm, and the Light Gradient Boosting Machine (LightGBM) algorithm.

The Decision Tree algorithm is a method for approximating the value of a discrete function (\citealt{1522531}; \citealt{article1}), and is a commonly used classification technique. It begins by processing the data and generating a set of readable rules and decision trees through an induction algorithm. These rules are then used to analyze new data, effectively classifying it. This algorithm is a popular method for predictive modeling in fields such as statistics, data mining, and machine learning (\citealt{Kotsiantis2013Decision}).
The Decision Tree algorithm is a tree-like structure that can be used to solve classification and regression problems. The construction process of a decision tree is based on the training set. Firstly, a feature is chosen as the root node, and then the dataset is divided into several subsets. Each subset of data has the same feature value. Then, a feature is selected as a node for classification in each data subset in turn. This process is repeated until a complete decision tree is generated. The advantage of a decision tree is that it is easy to understand and interpret, and the classification rules can be clearly understood. However, decision trees also suffer from the problem of overfitting, which needs to be optimized through methods such as pruning (\citealt{article1}). \cite{2011AJ....141..189V} explored 13 Decision tree algorithms for star/galaxy classification using SDSS DR7 data and proposed a novel method that can accurately distinguish between stars and galaxies with high precision.

The XGBoost algorithm, proposed by Tianqi Chen at the University of Washington, 
is a powerful tool for classification and regression tasks (\citealt{10.1145/2939672.2939785}; \citealt{Wang2020ASO}). 
It is composed of multiple Classification And Regression Tree decision trees, 
each of which learns the residual of the target value and the sum of all previous tree predictions. 
The final prediction is made by combining the results of all the trees (\citealt{article}). 
Though each weak classifier may not have a high global prediction accuracy, they can still have a very high prediction accuracy for specific aspects of the data(\citealt{10.1145/2939672.2939785}). 
By combining many classifiers with high local prediction accuracy, 
the XGBoost algorithm can achieve the effect of a strong classifier with high global prediction accuracy. 
It has gained popularity in data modeling competitions due to its excellent computing efficiency and prediction accuracy. 
It is widely used in various fields such as identifying stars, galaxies, and quasars from Beijing-Arizona Sky Survey (BASS) DR3 with accuracy more than 90$\%$ 
and classifying stars and galaxies using different models from SDSS DR7 (\citealt{2022MNRAS.509.2289L}; \citealt{2019ChA&A..43..539L}).

The LightGBM algorithm is a scalable machine learning system developed by Microsoft in 2017. 
It is an open-source project led by Guolin Ke, a winner of the first Alibaba Big Data Competition in 2014. 
The algorithm is based on the Gradient Boosting Decision Tree and is designed to reduce memory and computational requirements while also minimizing communication costs when used in parallel with multiple machines (\citealt{10.5555/3294996.3295074}). 
LightGBM can automatically identify effective data features and is considered an improved version of the XGBoost algorithm, providing faster training speed and lower memory consumption while maintaining a similar level of accuracy (\citealt{Wang2020ASO}). 
Due to its efficient performance on large-scale datasets, LightGBM has become a popular tool in data competitions. 
\citealt{2022MNRAS.513.5505M} used LightGBM to identify planets in simpler data and Transiting Exoplanet Survey Satellite (TESS) data, with very good results.

\subsection{ Data Balancing Algorithms }

During machine learning training, it has been observed that when the number of negative samples exceeds that of positive samples, the model focuses more on the characteristics of negative samples. Therefore, it is necessary to use some methods to reduce the issues caused by the serious imbalance between the number of positive and negative samples. To address this issue, we employed the Synthetic Minority Oversampling Technique (SMOTE) algorithm (\citealt{2018PaReL.103...32C}) and the Edited Nearest Neighbours (ENN) algorithm (\citealt{ENN1}) to balance the dataset. 

SMOTE is an algorithm used to address imbalanced classification problems. It generates synthetic samples of the minority class by interpolating between existing samples. The steps are: select a sample, identify its k nearest neighbors, randomly select one neighbor, generate a new synthetic sample, and repeat until the desired number of samples is reached. The new samples are combined with the original minority class samples to create a balanced dataset (\citealt{Chawla2002}). 

ENN is an undersampling technique commonly employed for addressing class imbalance by reducing the majority class to match the minority class (\citealt{ENN}). In order to determine whether the majority class of an observation's k nearest neighbors is the same as the class of the observation itself, the ENN  technique identifies the k nearest neighbors for each observation. If the majority class of an observation's k nearest neighbors differs from the class of the observation, both the observation and its k nearest neighbor are removed from the dataset (\citealt{ENN2}).

SMOTE and ENN algorithms were applied to the training data, resulting in the construction of 12 machine learning models. To compare with the generated 12 models and accurately identify symbiotic stars, we constructed a new training set with a nearly 1:1 ratio of positive and negative samples. We randomly selected 200 non-symbiotic stars from the training set and combined them with our 198 symbiotic stars to create the new training set. We applied the Decision Tree, XGBoost, and LightGBM algorithms to the new training set, resulting in the construction of a total of 16 machine learning models throughout our research process. Table \ref{tab1} shows the algorithms and constructions used for the 16 models. 

\subsection{ Results }
We randomly partitioned the dataset into a training set consisting of 80$\% $ of the data set and a test set consisting of 20$\%$ of the data set, with the aim of ensuring a randomized partition. The training set was utilized for training the machine learning models, while the test set was employed for evaluating the performance of the machine learning models.
We trained the models with Scikit-learn. Scikit-learn is a Python-based open-source machine learning library that offers an extensive collection of tools and algorithms for data analysis, including classification, regression, clustering, dimensionality reduction, and model selection (\citealt{scikit-learn};\citealt{sklearn_api}). During the training process, we employ 5-fold cross-validation and GridSearchCV to optimize the model parameters. Cross-validation is a statistical analysis method commonly used to evaluate prediction performance and maximize data utilization (\citealt{7544814}). GridSearchCV is a traditional approach for hyperparameter optimization in machine learning (\citealt{muller2016introduction}).

To evaluate the performance of our machine learning models, we used four key evaluation metrics: Precision, Recall, F1-Score, and AUC. These metrics are defined as follows:
\begin{enumerate}
    \item Precision: The proportion of true positive samples among the predicted positive samples. It is calculated as TP / (TP + FP).
    \item Recall: The proportion of true positive samples that are correctly predicted as positive. It is calculated as TP / (TP + FN).
    \item F1-Score: The harmonic mean of precision and recall. It is calculated as 2 * (precision * recall) / (precision + recall).
    \item AUC: The area under the ROC curve, which can be used to measure the performance of a classification model. The AUC value ranges from 0 to 1, with a higher value indicating better model performance.
\end{enumerate}
Here, TP represents the number of samples predicted by the model as symbiotic stars and are actually symbiotic stars. FP represents the number of samples predicted by the model as symbiotic stars but are actually non-symbiotic stars. FN represents the number of samples predicted by the model as non-symbiotic stars but are actually symbiotic stars. The primary evaluation metric for our trained model in this study was Precision, as our goal was to accurately identify a maximum number of symbiotic stars. 

The performance of 16 models on the test set is summarized in Table \ref{tab2}. For a comprehensive understanding of the parameters of these models, please refer to Appendix \ref{app:A}. Additionally, Figure \ref{Fig3} presents the feature importance of the trained models. From Figure \ref{Fig3}, it can be observed that the feature importance of W2 is very high in the models that used decision trees (Models 1, 2, 5, 6, 9, 10, 13, and 14). The feature importance of other magnitudes is very low, indicating that decision tree models, although easy to construct and structurally simple, may not be stable enough. In the models that utilized XGBoost (Models 3, 7, 11, and 15), W2 and W1 have relatively high feature importance. For the models that employed LightGBM (Models 4, 8, 12, and 16), the feature importance of W4 and W3 is relatively high. The performance of the trained models was evaluated using a separate set of 72 symbiotic stars. Table \ref{tab3} provides a clear view of the predictions made by each model for these 72 symbiotic stars.

\begin{table}
\centering
\caption{Summarizing the 16 best performing machine learning models}
\label{tab1}
\setlength{\tabcolsep}{1em}
\begin{tabular}{cccccccccc}
\hline\noalign{\smallskip}
&Decision Tree\_gini & &  Decision Tree\_entropy & & XGBoost & & LightGBM\\
\hline\noalign{\smallskip}
Raw dataset &Model 1 & &  Model 2 & & Model 3 & & Model 4\\
SMOTE &Model 5 & &  Model 6 & & Model 7 & & Model 8\\
ENN&Model 9 & &  Model 10 & & Model 11 & & Model 12\\
200 non-symbiotic stars &Model 13 & &  Model 14 & & Model 15 & & Model  16\\
\hline\noalign{\smallskip}
\end{tabular}
\end{table}

\begin{table}
\centering
\caption{The performance of the 16 best models in predicting symbiotic stars on the test set, presented as Precision, Recall, F1-Score, and the AUC scores.}
\label{tab2}
\setlength{\tabcolsep}{1pt}
\begin{tabular}{ccccccccccccccccccc}
 \hline\noalign{\smallskip}
&model 1 & model 2 & model 3 & model 4 & model 5 & model 6 & model 7  &  model 8 \\
 \hline\noalign{\smallskip}
Precision & 0.94 & 0.94 &0.92 & 0.94  & 0.96 & 0.92 & 0.91  & 0.94 \\
Recall & 0.88 & 0.88 &  0.88 & 0.92 & 0.92 & 0.94 & 0.98  & 0.90 \\
F1-Score & 0.91 & 0.91 &  0.90 & 0.93 & 0.94 & 0.93 & 0.94  & 0.92  \\
AUC & 0.94 & 0.94 & 0.94 & 0.96 & 0.96 & 0.97 & 0.99  & 0.95 \\
 \hline\noalign{\smallskip}
 & model 9 & model 10 & model 11 & model 12 & model 13 & model 14  & model 15 & model 16 \\
 \hline\noalign{\smallskip}
 Precision &0.94 &  0.96 & 0.94 & 0.94  & 0.97& 0.97 & 0.97 & 1.00 \\
Recall & 0.96 &  0.87 & 0.92  & 0.92 & 0.97 & 0.97 & 0.97 & 0.97 \\
F1-Score & 0.95 &  0.91 & 0.93 & 0.93 & 0.97 & 0.97& 0.97 & 0.99  \\
AUC & 0.98 &  0.93 & 0.96 & 0.96 & 0.97& 0.97 & 0.97 & 0.99  \\
 \hline\noalign{\smallskip}
\end{tabular}
\end{table}

\begin{table}
\centering
\caption{The results presented in the table are the predictions made by the 16 trained models for 72 known symbiotic stars (SySts)}
\label{tab3}
\setlength{\tabcolsep}{1pt}
\begin{tabular}{ccccccccccccccc}
 \hline\noalign{\smallskip}
&model 1 & model 2 & model 3 & model 4 & model 5 & model 6 & model 7 & model 8 \\
 \hline\noalign{\smallskip}
 \^{y} = SySts & 60 & 59 & 63 & 62  & 59 & 62 & 66 & 63  \\
  \hline\noalign{\smallskip}
& model 9 & model 10 & model 11 & model 12 & model 13 & model 14 & model 15 & model 16\\
 \hline\noalign{\smallskip}
 \^{y} = SySts &  62 & 53 & 64 & 62 & 70 & 70 & 70 & 70 \\
 \hline\noalign{\smallskip}
\end{tabular}
\end{table}

\section{Identifying Symbiotic Stars }
\label{sect:Results}
The most reliable method for determining symbiotic stars is through spectroscopic analysis.
The earliest criteria for identifying symbiotic stars were proposed by \citet{1932PASP...44...56M}, stating that a symbiotic star is a binary system with a combined spectrum. 
As the study of symbiotic stars progressed, more refined criteria for identifying symbiotic stars were proposed, 
with the most current and widely accepted criteria being those proposed by \citet{2000A&AS..146..407B}. 
However, it is important to note that these criteria are only applicable to burning symbiotic stars, 
and accreting-only symbiotic stars may not show these spectral features unless an outburst occurs (\citealt{2021MNRAS.506.4151M}). 
In this study, we used the criteria proposed by \citet{2000A&AS..146..407B} to classify symbiotic stars. These criteria are as follows:

\begin{enumerate}
    \item resence spectral features of late-type giants such as $\rm TiO$, $\rm H_2O$, $\rm CO$, $\rm CN$ and $\rm VO$ bands as well as $\rm Ca$$\rm\uppercase\expandafter{\romannumeral1}$, $\rm Ca$$\rm\uppercase\expandafter{\romannumeral2}$, $\rm Fe$$\rm\uppercase\expandafter{\romannumeral1}$ and $\rm Na$$\rm\uppercase\expandafter{\romannumeral1}$ absorption lines.
    \item the detection of some typical emission lines, for instance $\rm H$$\rm\uppercase\expandafter{\romannumeral1}$ and $\rm He$$\rm\uppercase\expandafter{\romannumeral1}$. emission lines of ions with an ionization potential of at least 35 eV (e.g. [$\rm O$$\rm\uppercase\expandafter{\romannumeral3}$], [$\rm Fe$$\rm\uppercase\expandafter{\romannumeral7}$]$\lambda $$\lambda $5727,6087, [$\rm He$$\rm\uppercase\expandafter{\romannumeral2}$]$\lambda $4686).
    \item the presence of strong emission lines of $\rm O$$\rm\uppercase\expandafter{\romannumeral6}$$\lambda $$\lambda $6830,7088 (\citealt{1997A&A...327..191M}; \citealt{2000A&AS..146..407B}; \citealt{2019ApJS..240...21A}).
\end{enumerate}

\citet{2019MNRAS.483.5077A} proposed a new method for determining symbiotic stars using a mid-infrared criterion. 
This criterion was used to successfully discover five previously unknown symbiotic stars in the galaxy (\citealt{2021MNRAS.502.2513A}).

We aimed to classify symbiotic stars among a sample of 11,226,252 sources from LAMOST DR9 v1.0 using 16 machine learning models. We cross-matched the sources with the AllWISE and 2MASS catalogs at a radius of 6 arcsec, and obtained magnitude data for a total of 10,849,157 sources. The models were then utilized to identify symbiotic stars. In order to indentifying more reliable symbiotic star prediction, we employed a methodology that integrates the results of multiple models to verify the classification of a star as a symbiotic star. This ensures that the star is identified as a symbiotic star by all of the models used in our analysis.

A total of 11,709 sources were jointly identify as symbiotic stars by the 16 models.
These sources were then cross-matched with SDSS DR17 at a radius of 6 arcsec, and 15 of them were found to have spectral information. 
Among these 15 candidates, one was classified as a galaxy by both surveys, while another was identified as a quasar by SDSS and a galaxy by LAMOST, and 13 were found to have spectra similar to symbiotic stars. 
We have created a catalog of the relevant information for these sources and made it available on the website\footnote{https://doi.org/10.12149/101183}.
The spectra of 2 of the extent 13 sources are more in line with the criteria of \citet{2000A&AS..146..407B} for determining the spectra of symbiotic stars.
The spectra of the extent 11 sources are similar to the accreting-only symbiotic star V562Lyr. 

Here we discussed the spectra of the two sources that have been newly predicted as symbiotic stars. 

\subsection{V* V603 Ori}
\begin{figure}
  \centering
  \includegraphics[width=\textwidth, angle=0]{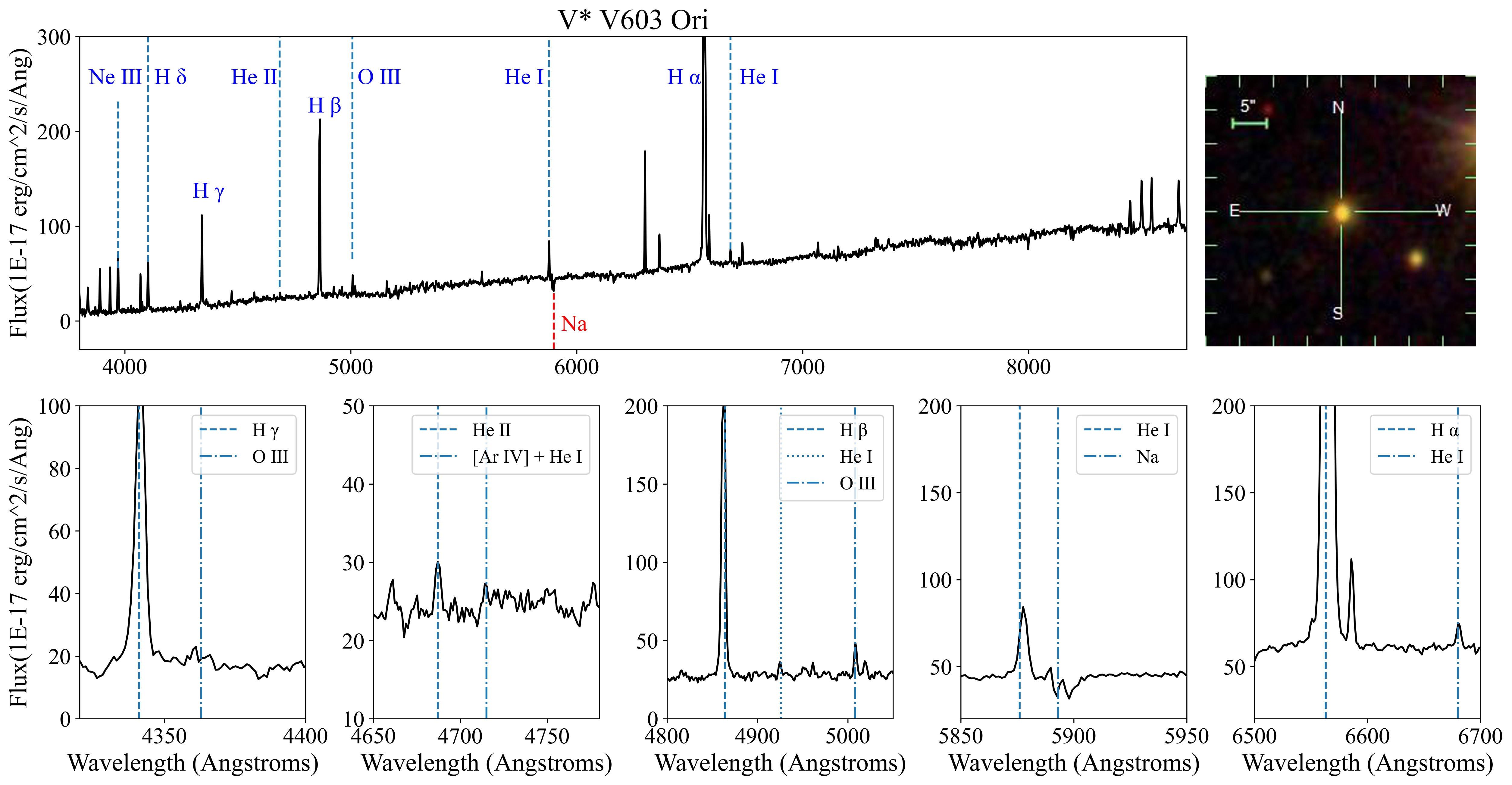}
  \caption{Low resolution spectra of V* V603 Ori from SDSS. Top left panel shows the V* V603 Ori observed spectra in SDSS. The top right panel displays the images of the V* V603 Ori in SDSS. South is down, west to the right. Bottom panels zoom in the $\rm H$$\gamma $ and $\rm O$$\rm\uppercase\expandafter{\romannumeral3}$ 4363{\AA}  lines, 
  $\rm He$$\rm\uppercase\expandafter{\romannumeral2}$ 4686{\AA}  line, $\rm H$$\beta $  and $\rm O$$\rm\uppercase\expandafter{\romannumeral3}$ 5007{\AA}  lines, $\rm H$$\alpha$ and $\rm He$$\rm\uppercase\expandafter{\romannumeral1}$ emission lines. }
  \label{Fig 4}
\end{figure}

\begin{figure}
  \centering
  \includegraphics[width=\textwidth, angle=0]{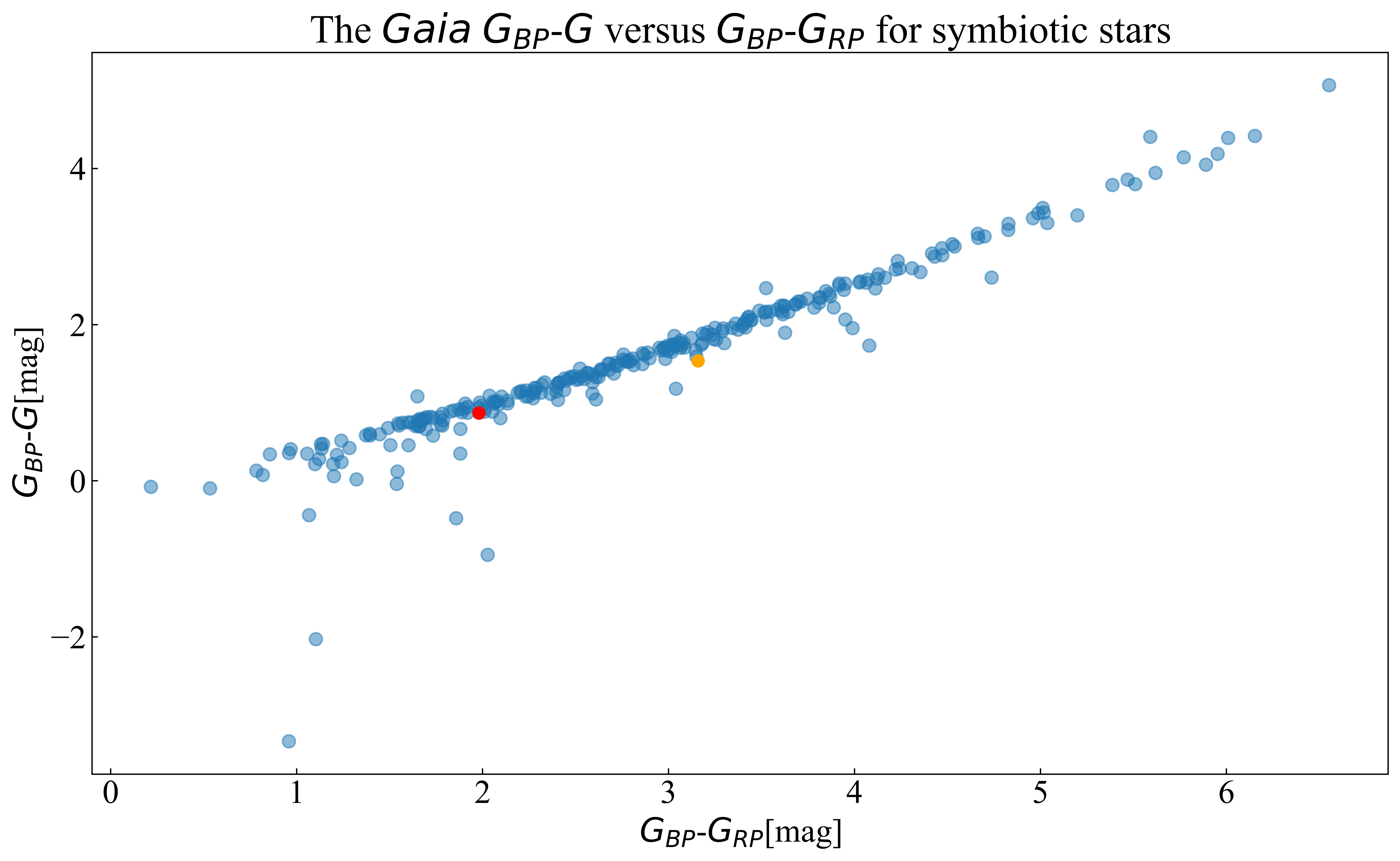}
  \caption{The $\textit{Gaia}$ $\mathit{G}_{\mathit{BP}}$-$\mathit{G}$ versus $\mathit{G}_{\mathit{BP}}$-$\mathit{G}_{\mathit{RP}}$ for symbiotic stars. The sources were identified through cross-matching 275 symbiotic stars with the Gaia catalog within a radius of 5 arcsec and are represented in blue. V* V603 Ori and V* GN Tau are shown in red and orange, respectively.  }
  \label{Fig5}
\end{figure}

V* V603 Ori (= 2MASS J05393983-0233160; RA2000 = 05 39 39.8292920832, DEC2000 = -02 33 16.040160936; see also Table \ref{tab4}) 
displays more prominent $\rm H$$\alpha$ , $\rm He$$\rm\uppercase\expandafter{\romannumeral1}$ lines, 
which increases the likelihood that it is a symbiotic star. 
Additionally, the presence of more distinct $\rm He$$\rm\uppercase\expandafter{\romannumeral2}$, $\rm H$$\beta$, $\rm He$$\rm\uppercase\expandafter{\romannumeral1}$ $\lambda$4930, $\rm O$$\rm\uppercase\expandafter{\romannumeral3}$ $\lambda$5007, and also $\rm H$$\gamma $(Figure \ref{Fig 4}) lines further supports this possibility. 
However, V* V603 Ori differs from the newly discovered symbiotic star DR2J141301.4-6533201.1 in \citet{2021MNRAS.502.2513A}, as the latter lacks the $\rm O$$\rm\uppercase\expandafter{\romannumeral3}$ $\lambda $4363 line, 
yet is still considered a symbiotic star. V* V603 Ori also exhibits some spectral characteristics of red giants, 
such as a molecular band containing $\rm TiO$ and absorption Na lines, as well as emission lines like$\rm Ne$$\rm\uppercase\expandafter{\romannumeral3}$. 
According to the magnitude information of AllWISE and 2MASS of V* V603 Ori, it is classified as an S+IR-type symbiotic star. 
However, the TESS survey has not captured any obvious light variation in V* V603 Ori, thus it cannot be confirmed as a symbiotic star through light variation analysis. 
Fortunately, the renormalised unit weight error (RUWE) parameter in GaiaDR3 is an essential indicator for determining whether a star is binary. 
According to \citet{2021gdr3.reptE..13H}, when the RUWE value is 1.4, it is likely that the star is not single. In the case of V* V603 Ori, the RUWE value is 1.4. 
\citet{2017A&A...606A.110I} explored various other emission lines including [N II], [O III], [Ne III] and HeI lines, in order to distinguish planetary nebulae from symbiotic stars and proposed criteria for symbiotic stars based on emission line ratios. for example, $\log ([\mathrm{O} \text { III] } 5006 /[\mathrm{N} \text { II] } 5755)<2.6 \text { and } \log ([\mathrm{Ne} \text { III] } 3869 /[\mathrm{O} \text { III }] 4363)<0.45$. V* V603 Ori meets this criterion. V* V603 Ori has been classified as an M0 spectral type star by LAMOST.
The symbiotic stars were cross-matched with the Gaia catalog within 5 arcsec , and corresponding diagnostic colour-colour diagrams were plotted for The Gaia $\mathit{G}_{\mathit{BP}}$-$\mathit{G}$ versus $\mathit{G}_{\mathit{BP}}$-$\mathit{G}_{\mathit{RP}}$. In the figure\ref{Fig5}, V* V603 Ori was highlighted in red.
Based on the above information, we classify V* V603 Ori as a  newly discovered symbiotic star.

\begin{table}\footnotesize\centering
\begin{center}
\caption[]{Basic properties of V* V603 Ori and V* GN Tau. Data are from Gaia DR3 (\citealt{2022arXiv220800211G}), 2MASS (\citealt{2006AJ....131.1163S}), and AllWISE (\citealt{2010AJ....140.1868W}).}
\label{tab4}

 \begin{tabular}{cccccccccc}
  \hline\noalign{\smallskip}
Soucre & RA  & DEC &   \textit {J} &  \textit {H}    &\textit {Ks}   &\textit {W}1   &\textit {W}2   & \textit {W}3  &\textit {W}4   \\
 &(J2000) & (J2000)& (mag) & (mag) & (mag) & (mag) & (mag) & (mag) & (mag) \\
  \hline\noalign{\smallskip}
  V* V603 Ori & 84.915955 &-02.554456       & 12.22 ± 0.03     & 10.96 ± 0.02    & 10.07 ± 0.02      & 8.96 ± 0.02    & 8.33 ± 0.03 	    & 6.76 ± 0.02 	    & 5.05 ± 0.04 \\
  V* GN Tau & 69.837176 & +25.750563      		 & 10.20 ± 0.03   		 & 8.89 ± 0.03   		 & 8.06 ± 0.03  		 & 7.16 ± 0.04   		 & 6.53 ± 0.02  		 & 4.91 ± 0.01   & 3.05 ± 0.02  \\
  \noalign{\smallskip}\hline
\end{tabular}
\end{center}
\end{table}

\subsection{V* GN Tau}
\begin{figure}
  \centering
  \includegraphics[width=0.99\textwidth, angle=0]{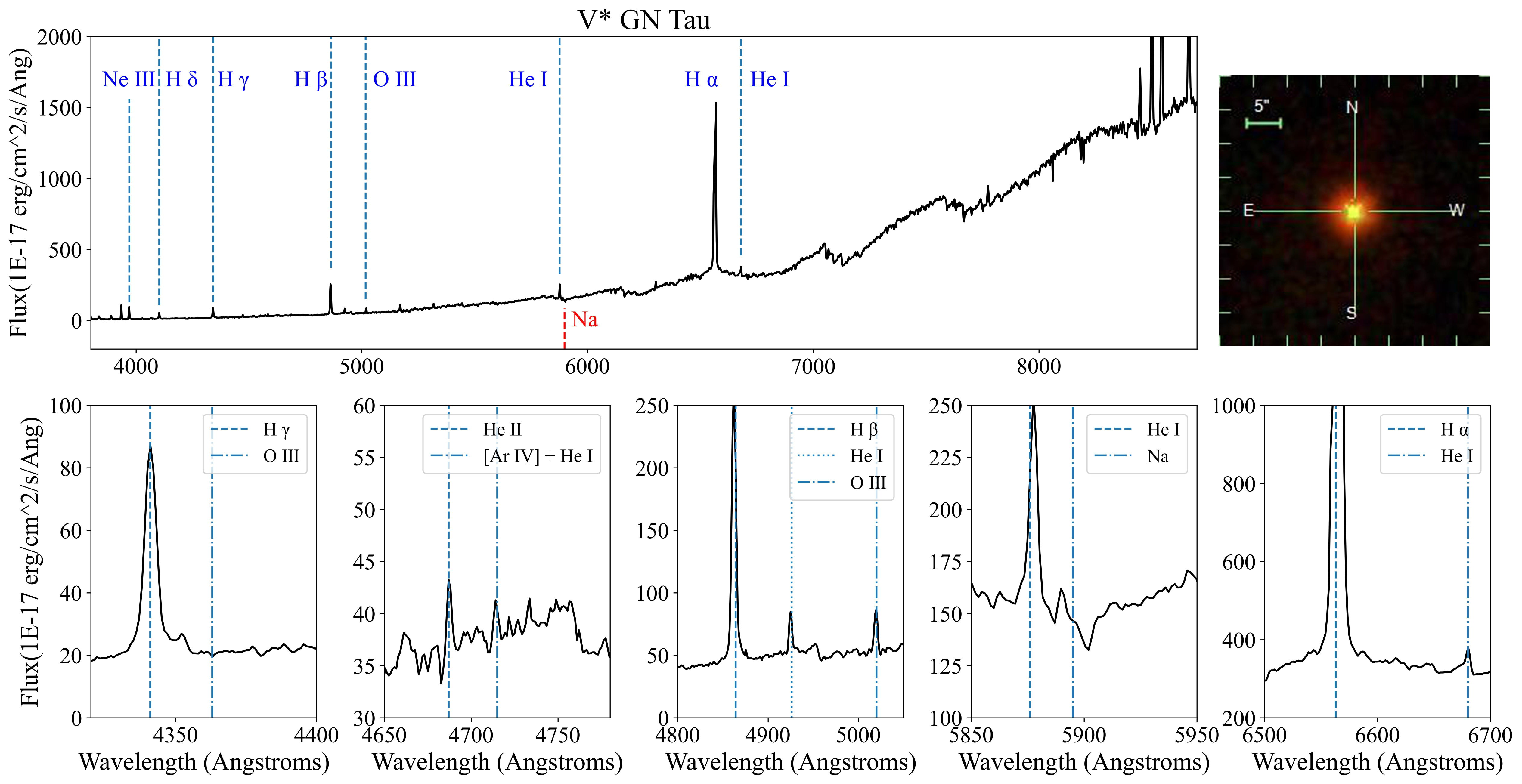}
  \caption{Low resolution spectra of V* GN Tau Ori from SDSS. Top left panel shows the V* GN Tau observed spectra in SDSS. The top right panel displays the images of the V* GN Tau in SDSS. South is down, west to the right. Bottom panels zoom in the $\rm H$$\gamma $ and $\rm O$$\rm\uppercase\expandafter{\romannumeral3}$ 4363{\AA}  lines, 
  $\rm He$$\rm\uppercase\expandafter{\romannumeral2}$ 4686{\AA}  line, $\rm H$$\beta $  and $\rm O$$\rm\uppercase\expandafter{\romannumeral3}$ 5007{\AA}  lines, $\rm H$$\alpha$ and $\rm He$$\rm\uppercase\expandafter{\romannumeral1}$ emission lines. }
  \label{Fig6}
\end{figure}

The source J043920.90+254502.1 in LAMOST cross-matched with AllWISE is V* GN Tau(see also Table \ref{tab4}), which is predicted to be a symbiotic star by our model. 
The SDSS spectrogram of V* GN Tau displays prominent $\rm H$$\alpha$ and $\rm He$$\rm\uppercase\expandafter{\romannumeral1}$ lines (Figure \ref{Fig6}), suggesting that it is a strong candidate for being a symbiotic star. 
Other noticeable lines in the spectrum include $\rm He$$\rm\uppercase\expandafter{\romannumeral2}$ ,$\rm H$$\beta$, $\lambda$4930, $\rm O$$\rm\uppercase\expandafter{\romannumeral3}$ $\lambda$5007, and also $\rm H$$\gamma $ and $\rm Na$$\rm\uppercase\expandafter{\romannumeral1}$ absorption, as well as a molecular band of $\rm TiO$. 
Although there is a lack of $\rm O$$\rm\uppercase\expandafter{\romannumeral3}$ $\lambda $4363 lines, it does not necessarily exclude the possibility of V* GN Tau being a symbiotic star. 
Other known symbiotic stars such as AS 201, V3811 Sgr, and V407 Cyg also display a similar absence of certain spectral lines (refer to Table A.7 in \citet{2009syst.book.....K}). 
The spectrum of V* GN Tau is similar to that of DR2J175346.2-284826.16, which was previously identified as a symbiotic candidate by \citet{2021MNRAS.502.2513A}.

The magnitude information of V* GN Tau in AllWISE and 2MASS supports its classification as an S+IR-type symbiotic star according to Akras' mid-infrared color standards. 
However, the lack of clear light variation in V* GN Tau as captured by the TESS survey hinders us from fully confirming its status as a symbiotic star through photometric analysis. 
The RUWE parameter value of V* GN Tau is not available in GaiaDR3, thus it is not possible to determine whether V* GN Tau is a binary star.  V* GN Tau also meets  \citet{2017A&A...606A.110I} criterion. V* GN Tau has been classified as an M4/3 spectral type star by LAMOST. In the figure\ref{Fig5}, V* GN Tau was highlighted in orange.
But V* GN Tau is still identified as a symbiotic star based on its spectral information.

\section{Discussion and conclusions}
\label{sect:D and C}
In this study, the AllWISE and 2MASS magnitude data of known symbiotic stars were used to identify potential symbiotic stars through a machine learning approach. 
The constructed model processed 10,849,157 AllWISE and 2MASS catalogs from LAMOST DR9 sources and identified 11,709 symbiotic stars. 

We analyzed the SDSS spectra of these symbiotic star candidates and discovered two new symbiotic stars, namely V* V603 Ori and V* GN Tau. The spectra of these two symbiotic stars were found to be consistent with the spectral classification criteria of  \citet{2000A&AS..146..407B} and the mid-infrared color classification criteria of \citet{2019MNRAS.483.5077A}.

Actually, the criteria for identifying symbiotic stars based on their spectra are typically used for active, or ``burning" symbiotic stars. 
However, some symbiotic stars can remain hidden for a long period of time, referred to as accreting-only symbiotic stars (\citealt{2022arXiv221113193P}). 
These stars do not exhibit the spectroscopic characteristics of active symbiotic stars and do not have distinct emission lines in the optical band, 
but instead, are bright in the ultraviolet spectrum (\citealt{2013A&A...559A...6L}; \citealt{2016MNRAS.461L...1M}; \citealt{2021MNRAS.505.6121M}; \citealt{2021MNRAS.506.4151M}). 
An example of this is V1988 Sgr, whose spectrum was analyzed by \citet{2021MNRAS.506.4151M} and found to not match the spectral features of active symbiotic stars. 
Despite this, \citet{2021MNRAS.506.4151M} concluded that V1988 Sgr could not be ruled out as an accreting-only symbiotic star. 
Another example is V562 Lyr, which was identified as a symbiotic star despite the absence of prominent $\rm H$ and $\rm He$ lines. 
This weakening or disappearance of optical emission lines in the spectra of symbiotic stars has been found to be due to a decrease in the accretion rate by \citet{2022arXiv221113193P}.

Among the 11 remaining stars in our candidates catalog, we observed similar spectra, tentatively classifying them as accreting-only symbiotic star candidates. The next step in our research will be to observe these additional 11 candidates and gather more evidence to make a definitive classification. Further observations and data are necessary to confirm this classification.

\begin{acknowledgements}
This work received the generous support of  the Natural Science Foundation of Xinjiang No.2021D01C075, the National Natural Science Foundation of China, project Nos. 12163005, 12003025, U2031204 and 11863005, the science research grants from the China Manned Space Project with NO. CMS-CSST-2021-A10, the Scientific Research Program of the Higher Education Institution of Xinjiang (No. XJEDU2022P003).
Data resources are supported by China National Astronomical Data Center (NADC) and Chinese Virtual Observatory (China-VO). This work is supported by Astronomical Big Data Joint Research Center, co-founded by National Astronomical Observatories, Chinese Academy of Sciences and Alibaba Cloud.
This work made use of Astropy:\footnote{http://www.astropy.org} a community-developed core Python package and an ecosystem of tools and resources for astronomy (\citealt{2013A&A...558A..33A}; \citealt{2018AJ....156..123A}; \citealt{2022ApJ...935..167A}).
This publication makes use of data products from the Wide-field Infrared Survey Explorer, which is a joint project of the University of California, Los Angeles, and the Jet Propulsion Laboratory/California Institute of Technology, and NEOWISE, which is a project of the Jet Propulsion Laboratory/California Institute of Technology. WISE and NEOWISE are funded by the National Aeronautics and Space Administration.
This publication makes use of data products from the Two Micron All Sky Survey, which is a joint project of the University of Massachusetts and the Infrared Processing and Analysis Center/California Institute of Technology, funded by the National Aeronautics and Space Administration and the National Science Foundation.
Guoshoujing Telescope (the Large Sky Area Multi-Object Fiber Spectroscopic Telescope LAMOST) is a National Major Scientific Project built by the Chinese Academy of Sciences. Funding for the project has been provided by the National Development and Reform Commission. LAMOST is operated and managed by the National Astronomical Observatories, Chinese Academy of Sciences.
Funding for the Sloan Digital Sky 
Survey IV has been provided by the 
Alfred P. Sloan Foundation, the U.S. 
Department of Energy Office of 
Science, and the Participating 
Institutions. 
SDSS-IV acknowledges support and 
resources from the Center for High 
Performance Computing  at the 
University of Utah. The SDSS 
website is www.sdss4.org.
SDSS-IV is managed by the 
Astrophysical Research Consortium 
for the Participating Institutions 
of the SDSS Collaboration including 
the Brazilian Participation Group, 
the Carnegie Institution for Science, 
Carnegie Mellon University, Center for 
Astrophysics | Harvard \& 
Smithsonian, the Chilean Participation 
Group, the French Participation Group, 
Instituto de Astrof\'isica de 
Canarias, The Johns Hopkins 
University, Kavli Institute for the 
Physics and Mathematics of the 
Universe (IPMU) / University of 
Tokyo, the Korean Participation Group, 
Lawrence Berkeley National Laboratory, 
Leibniz Institut f\"ur Astrophysik 
Potsdam (AIP),  Max-Planck-Institut 
f\"ur Astronomie (MPIA Heidelberg), 
Max-Planck-Institut f\"ur 
Astrophysik (MPA Garching), 
Max-Planck-Institut f\"ur 
Extraterrestrische Physik (MPE), 
National Astronomical Observatories of 
China, New Mexico State University, 
New York University, University of 
Notre Dame, Observat\'ario 
Nacional / MCTI, The Ohio State 
University, Pennsylvania State 
University, Shanghai 
Astronomical Observatory, United 
Kingdom Participation Group, 
Universidad Nacional Aut\'onoma 
de M\'exico, University of Arizona, 
University of Colorado Boulder, 
University of Oxford, University of 
Portsmouth, University of Utah, 
University of Virginia, University 
of Washington, University of 
Wisconsin, Vanderbilt University, 
and Yale University.
\end{acknowledgements}

\bibliographystyle{raa}
\bibliography{bibtex}

\label{lastpage}

\appendix
\section{Algorithm and Parameter Details of the 16 Machine Learning Models}
\label{app:A}
 Here is a detailed description of the parameters of these 16 models:

Models 1 through 4 were trained without using the SMOTE and ENN algorithms for data balancing. The training set consisted of 146 symbiotic stars and 20,139 non-symbiotic stars. Model 1 was a Decision Tree model that used the Gini index as the splitting criterion. Model 2 was a Decision Tree model that used entropy as the splitting criterion. Model 3 was an XGBoost model, and Model 4 was a LightGBM model. 

Models 5 through 8 were trained on a balanced dataset using the SMOTE algorithm. We conducted multiple rounds of parameter selection and determined the optimal parameter values for SMOTE algorithm using grid search, which are k\_neighbors = 1 and sampling\_strategy = 'minority'. After applying SMOTE, the training set consisted of 20,139 symbiotic stars and 20,139 non-symbiotic stars. Model 5 was a DecisionTree model trained on the SMOTE-balanced data using the Gini index as the splitting criterion, similar to Model 1. Model 6 was a DecisionTree model trained on the SMOTE-balanced data using the entropy criterion as the splitting criterion, similar to Model 2. Model 7 was an XGBoost model trained on the SMOTE-balanced data. Model 8 was a LightGBM model trained on the SMOTE-balanced data.

Models 9 through 12 were trained using the ENN algorithm to balance the data. We conducted multiple rounds of parameter selection and determined the optimal parameter values for ENN algorithm using grid search, which are kind\_sel = 'mode', n\_neighbors = 1 and sampling\_strategy = 'majority'. After the application of ENN, the training set consisted of 20,120 symbiotic stars and 146 non-symbiotic stars. Model 9 is a DecisionTree model trained on the ENN-balanced data using the Gini index as the splitting criterion, similar to Model 1. Model 10 is a DecisionTree model trained on the ENN-balanced data using the entropy criterion as the splitting criterion, similar to Model 2. Model 11 is an XGBoost model trained on the ENN-balanced data. Model 12 is a LightGBM model trained on the ENN-balanced data. 

Models 13 through 16 were trained without using the SMOTE and ENN algorithms for data balancing. The training set consisted of 159 symbiotic stars and 159 non-symbiotic stars. Model 13 was a Decision Tree model that used the Gini index as the splitting criterion. Model 14 was a Decision Tree model that used entropy as the splitting criterion. Model 15 was an XGBoost model, and Model 16 was a LightGBM model.

The optimal model parameters obtained after training are as follows:

\setlength{\parindent}{0pt} 
Model 1: 
\begin{minipage}[t]{\linewidth}
class\_weight = None, criterion = 'gini', max\_depth = 7, max\_features = 6, min\_samples\_leaf = 5, min\_samples\_split = 51, splitter = 'best'.
\end{minipage}

\setlength{\parindent}{0pt} 
Model 2: 
\begin{minipage}[t]{\linewidth}
class\_weight = None, criterion = 'entropy', max\_depth = 10, max\_features = 3, min\_samples\_leaf = 2, min\_samples\_split = 3, splitter = 'best'.
\end{minipage}

\setlength{\parindent}{0pt} 
Model 3: 
\begin{minipage}[t]{\linewidth}
colsample\_bytree = 0.6, gamma = 0, learning\_rate = 0.1, max\_depth = 15, min\_child\_weight = 1, n\_estimators = 580, reg\_alpha = 0.1, reg\_lambda = 0.1, scale\_pos\_weight = 300, subsample = 0.9.
\end{minipage}

\setlength{\parindent}{0pt} 
Model 4: 
\begin{minipage}[t]{\linewidth}
colsample\_bytree = 0.6, learning\_rate = 0.1, max\_depth = 11, min\_child\_weight = 1, n\_estimators = 500, reg\_alpha = 0, reg\_lambda = 0, scale\_pos\_weight = 300, subsample = 0.9.
\end{minipage}

\setlength{\parindent}{0pt} 
Model 5: 
\begin{minipage}[t]{\linewidth}
class\_weight = None, criterion = 'gini', max\_depth = 15, max\_features = 5, min\_samples\_leaf = 2, min\_samples\_split = 6, splitter = 'best'.
\end{minipage}

\setlength{\parindent}{0pt} 
Model 6: 
\begin{minipage}[t]{\linewidth}
class\_weight = None, criterion = 'entropy', max\_depth = 15, max\_features = 7, min\_samples\_leaf = 2, min\_samples\_split = 6, splitter = 'best'.
\end{minipage}

\setlength{\parindent}{0pt} 
Model 7: 
\begin{minipage}[t]{\linewidth}
colsample\_bytree = 0.6, gamma = 0.1, learning\_rate = 0.1, max\_depth = 15, min\_child\_weight = 0.1, n\_estimators = 100, reg\_alpha = 0, reg\_lambda = 0, scale\_pos\_weight = 1, subsample = 0.8.
\end{minipage}

\setlength{\parindent}{0pt} 
Model 8: 
\begin{minipage}[t]{\linewidth}
colsample\_bytree = 0.6, learning\_rate = 0.1, max\_depth = 11, min\_child\_weight = 1, n\_estimators = 500, reg\_alpha = 0, reg\_lambda = 0, scale\_pos\_weight = 1, subsample = 0.6.
\end{minipage}

\setlength{\parindent}{0pt} 
Model 9: 
\begin{minipage}[t]{\linewidth}
class\_weight = None, criterion = 'gini', max\_depth = 15, max\_features = 4, min\_samples\_leaf = 3, min\_samples\_split = 6, splitter = 'best'.
\end{minipage}

\setlength{\parindent}{0pt} 
Model 10: 
\begin{minipage}[t]{\linewidth}
class\_weight = None, criterion = 'entropy', max\_depth = 20, max\_features = 7, min\_samples\_leaf = 9, min\_samples\_split = 15, splitter = 'best'.
\end{minipage}

\setlength{\parindent}{0pt} 
Model 11: 
\begin{minipage}[t]{\linewidth}
 colsample\_bytree = 0.6, gamma = 0.1, learning\_rate = 0.1, max\_depth = 15, min\_child\_weight = 0.1, n\_estimators = 100, reg\_alpha = 0, reg\_lambda = 0, scale\_pos\_weight = 200, subsample = 0.8.
\end{minipage}

\setlength{\parindent}{0pt} 
Model 12: 
\begin{minipage}[t]{\linewidth}
 colsampe\_bytree = 0.6, learning\_rate = 0.1, max\_depth = 11, min\_child\_weight = 1, n\_estimators = 500, reg\_alpha = 0, reg\_lambda = 0, scale\_pos\_weight = 300, subsample = 0.6.
\end{minipage}

\setlength{\parindent}{0pt} 
Model 13: 
\begin{minipage}[t]{\linewidth}
class\_weight = None, criterion = 'gini', max\_depth = 5, max\_features = 4, min\_samples\_leaf = 5, min\_samples\_split = 51, splitter = 'best'.
\end{minipage}

\setlength{\parindent}{0pt} 
Model 14: 
\begin{minipage}[t]{\linewidth}
class\_weight = None, criterion = 'entropy', max\_depth = 10, max\_features = 5, min\_samples\_leaf = 5, min\_samples\_split = 5, splitter = 'best'.
\end{minipage}

\setlength{\parindent}{0pt} 
Model 15: 
\begin{minipage}[t]{\linewidth}
colsample\_bytree = 0.6, gamma = 1, learning\_rate = 0.1, max\_depth = 15, min\_child\_weight = 1, n\_estimators = 70, reg\_alpha = 1, reg\_lambda = 1, scale\_pos\_weight = 1, subsample = 0.9.
\end{minipage}

\setlength{\parindent}{0pt} 
Model 16: 
\begin{minipage}[t]{\linewidth}
colsample\_bytree = 0.6, learning\_rate = 0.1, max\_depth = 10, min\_child\_weight = 1, n\_estimators = 50, reg\_alpha = 0, reg\_lambda = 0, scale\_pos\_weight = 1, subsample = 0.9.
\end{minipage}

\end{document}